\documentclass[twocolumn,aps,prl,english,showpacs,preprintnumbers,amsmath,amssymb]{revtex4}
\usepackage[latin1]{inputenc}
\usepackage{amsmath}
\usepackage{babel}
\usepackage{graphics}
\usepackage{amssymb}
\usepackage[usenames,dvipsnames]{color}
\usepackage{soul}

\usepackage{bm}

\def\ave#1{\langle #1\rangle}

\newcommand{\up}{\uparrow}
\newcommand{\dn}{\downarrow} 

\newcommand{\iv}{\mathbf{i}}
\newcommand{\jv}{\mathbf{j}}
\newcommand{\kv}{\mathbf{k}}

\newcommand{\bdd}{\boldsymbol\delta}
\newcommand{\bdp}{\boldsymbol\varphi}

\begin{document}


\title{Size and shape of Mott regions for fermionic atoms in a two-dimensional optical lattice}

\author{Tiago  \surname{Mendes-Santos}, Thereza  \surname{Paiva}, and Raimundo R. \surname{dos Santos}} 

\affiliation{Instituto de F\'\i sica, 
                 Universidade Federal do Rio de Janeiro, 
                 Caixa Postal 68528,
                 21941-972 Rio de Janeiro RJ, 
                 Brazil}

\begin{abstract}
We investigate the harmonic-trap control of size and shape of Mott regions in the Fermi Hubbard model on a square optical lattice. 
The use of Lanczos diagonalization on clusters with twisted boundary conditions, followed by an average over 50-80 samples, drastically reduce finite-size effects in some ground state properties; 
calculations in the grand canonical ensemble together with a local-density approximation (LDA) allow us to simulate the radial density distribution. 
We have found that as the trap closes, the atomic cloud goes from a metallic state, to a Mott core, and to a Mott ring; the coverage of Mott atoms reaches a maximum at the core-ring transition. 
A `phase diagram' in terms of an effective density and the on-site repulsion is proposed, as a guide to maximize the Mott coverage.
We also predict that the usual experimentally accessible quantities, the global compressibility and the average double occupancy (rather, its density derivative) display detectable signatures of the core-ring transition.
Some spin correlation functions are also calculated, and predict the existence N\'eel ordering within Mott cores and rings.
\end{abstract}

\date{ \today}

\pacs{
	03.75.Ss 	
	05.30.Fk   
         67.85.-d 	
         71.10.Fd 	 
         71.30.+h 	 
}

\maketitle  

Ultracold atoms in optical lattices have emerged as an invaluable tool in the study of strongly correlated fermions, due to the increasing control and manipulation achieved in experiments \cite{DeMarco99,Bloch08,Esslinger10}. 
On the one hand, the range of applications has broadened, shedding light into the problems of coherence and entanglement \cite{Bloch08}; also, imaging has become an important asset in the study of ultracold atoms \cite{Sherson10}, so that visual analyses of systems undergoing quantum phase transitions can provide new insights into dynamical properties \cite{Simon11}.
On the other hand, ultracold atoms in optical lattices may act as quantum simulators, in the sense that one can devise experiments to extract information on many-body models which have so far eluded a wide range of theoretical approaches \cite{Esslinger10}. 
For instance, the compressibility of $^{40}$K atoms in an optical lattice, has been measured as a function of both trap compression and on-site repulsion \cite{Schneider08}: 
as the trap is deepened, or the repulsion between atoms is increased, first a Mott `shell' develops within an overall metallic phase, followed by the emergence of a Mott core; the appearance of these Mott regions is in line with theoretical predictions for a trapped Hubbard model \cite{Helmes08,Fuchs11,Paiva11,Chiesa11}.
The controlled appearance of Mott regions represents an important step towards the experimental observation of antiferromagnetism in fermionic atomic clouds \cite{Duarte14}; 
as such it could also help elucidating its role as a precursor to the superconducting state in cuprates \cite{Esslinger10}, especially in relation with the role played by the two-dimensional CuO$_2$ planes. 
In this respect, one must be able to characterize the size and form of the Mott region for the trap parameters at hand.
With this in mind, here we focus on aspects of the latter issue for the case of a square optical lattice; this choice of lattice is motivated both by the possibility of singling out some special features brought about by the van Hove singularity, and by the use of an unbiased calculational method (see below).
Our main result is that the use of experimentally accessible \emph{global} quantities, such as compressibility and (derivative of) average double occupancy, can be used to map out \emph{local} phases and phase separation; accordingly, we have established a `phase diagram' describing the boundaries between Mott ring and core in terms of trap depth, fermion repulsion, and number of particles.

\begin{figure}[t]
{\centering\resizebox*{8.0cm}{!}{\includegraphics*{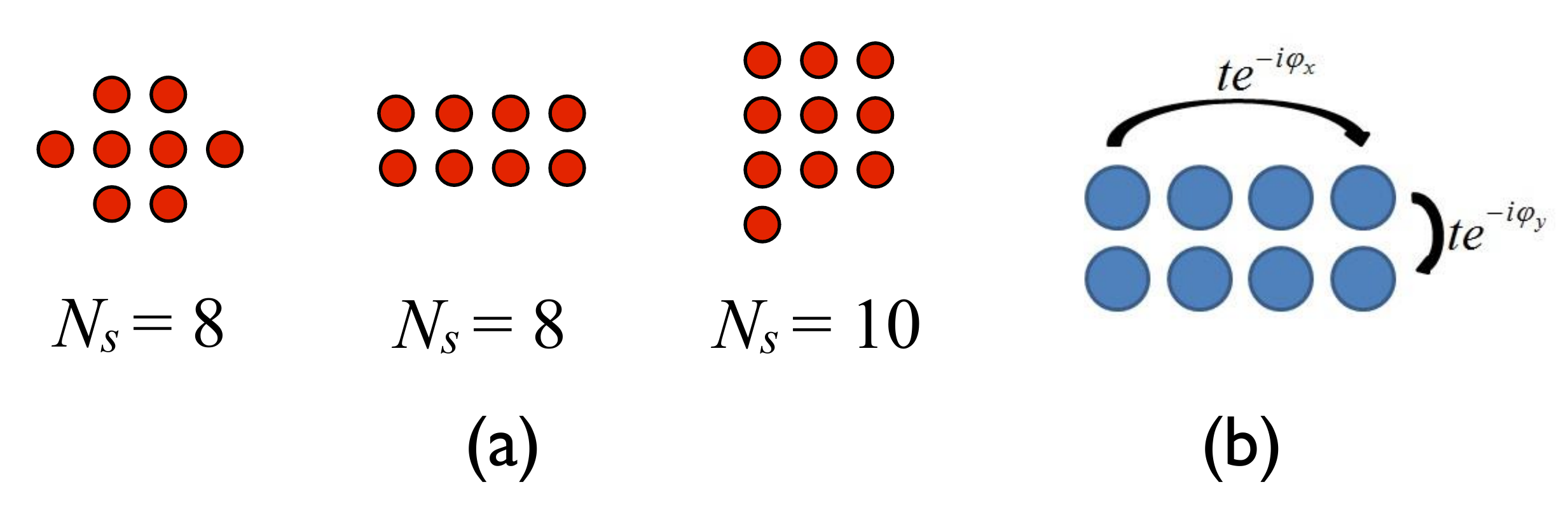}}}
\caption{(Color online) (a) Square lattice clusters; (b) `Twisted' boundary conditions are introduced through phases $\varphi_x$ and $\varphi_y$ in the hopping term (see text).
}
\vskip -0.4cm
\label{fig:clusters} 
\end{figure}

We assume the trap conditions are such that the 
fermionic atoms are described by a repulsive Hubbard model with a position-dependent chemical potential (due to a parabolic confining potential), namely,
\begin{align}
\mathcal{H}= &- \sum_{\ave{\iv , \jv},\sigma} t_{\iv\jv} \left( c_{\iv\sigma}^\dagger c_{\jv\sigma}^{\phantom{\dagger}} \right)
+ U \sum_\iv n_{\iv\up} n_{\iv\dn}\nonumber\\
&-    \sum_{\iv} (\mu_0-V_t r_\iv^2) (n_{\iv\up} + n_{\iv\dn}),
\end{align}
where, in standard notation, $\iv$ runs over the $N_s$ sites of the lattice, $c_{\iv\, \sigma}^\dagger$ ($c_{\iv\, \sigma}^{\phantom{\dagger}}$) creates (annihilates) a fermion at site $\iv$ in the spin state $\sigma=\,\up$ or $\dn$, and $n_{\iv\sigma}=c_{\iv\, \sigma}^\dagger c_{\iv\, \sigma}^{\phantom{\dagger}}$; $t_{\iv\jv}$ is the hopping integral (or tunnelling rate) between sites $\iv$ and $\jv$, $U$ is the magnitude of the on-site repulsion, $\mu_0$ is the (bare) chemical potential, $V_t$ measures the trap opening, and $r_\iv$ measures the distance of site $\iv$ to the center of the trap.

\begin{figure}[t]
{\centering\resizebox*{8.5cm}{!}{\includegraphics*{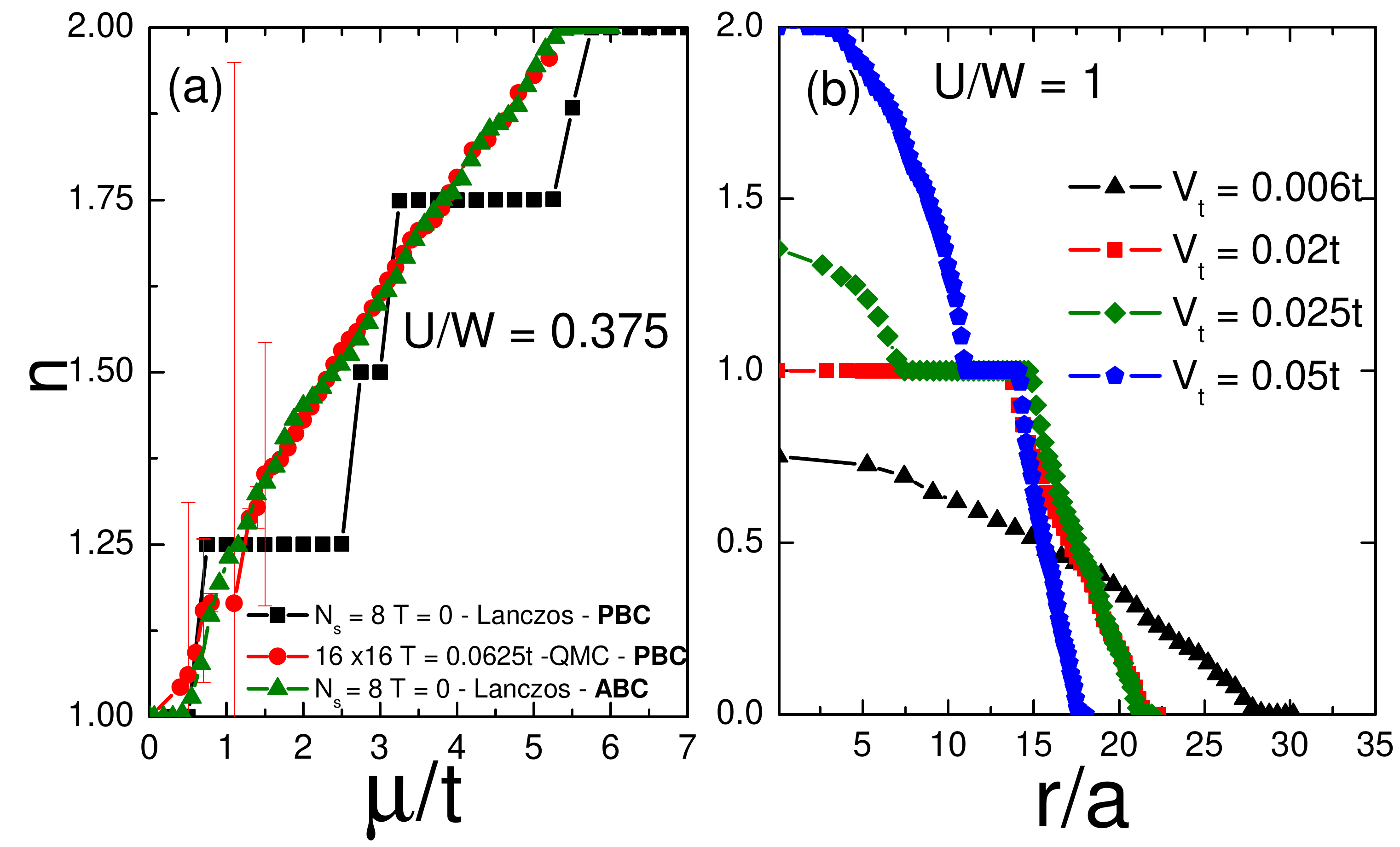}}}
\caption{(Color online) (a) Fermion density as a function of the shifted chemical potential (i.e., $\mu\to\mu-U/2$), obtained by different methods: Lanczos diagonalization with periodic (PBC) and averaged boundary conditions (ABC), and Quantum Monte Carlo at a finite temperature. 
(b) Radial distribution for $N=10^3$ atoms, for different trap depths $V_t$. 
The value of $U$ is different in both panels.
}
\vskip -0.6cm
\label{fig:platx} 
\end{figure}

We start by considering the homogeneous Hamiltonian (i.e., $V_t=0$) on the clusters shown in Fig.\,\ref{fig:clusters}(a), the idea being that the whole square lattice can be generated by suitable translations of the clusters. 
Therefore, whenever a fermion hops between two adjacent copies of the cluster [or, equivalently, reenters through an opposite edge of the same cluster, as in Fig.\,\ref{fig:clusters}(b)], the hopping term picks up a phase, $t_{\iv\jv}= te^{i\varphi_{\iv\jv}}$, where, in this context, $\iv$ and $\jv$ refer to border sites; otherwise, $t_{\iv\jv}= t$ (from now on, the bandwidth $W=8t$ sets the energy scale). 
Periodic boundary conditions (BC's) correspond to $\varphi_{\iv\jv}=0,\forall\, \iv , \jv$, and antiperiodic BC's to $\varphi_{\iv\jv}=\pi, \forall\, \iv, \jv$. 
Different sets of pairs $\bdp\equiv(\varphi_x,\varphi_y)$ yield different allowed $\kv$-points, so that 
finite-size effects can be minimised by considering an ensemble of \emph{random} 
sets $\bdp^{(\ell)},\,\ell=1,\ldots,M$; for each $\bdp^{(\ell)}$, we calculate the quantities of interest (see below), and then perform an average over the $M$ realizations \cite{Loh88,Gammel93,Gros96,Lin01}.
In so doing, the number of allowed $\kv$-points increases, mimicking a dense Brillouin Zone (BZ) similar to that occurring for $N_s\to\infty$.
Clusters up to 10 sites are particularly amenable to Lanczos diagonalizations, and 
Figure \ref{fig:platx}(a) compares three different results for the fermion density as a function of the chemical potential.
While Lanczos diagonalization with periodic BC's displays spurious \emph{plateaux}  (due to `closed shell' effects; see, e.g., Ref.\,\cite{Mondaini12}), the use of averaged BC's wipes out these effects, while preserving the Mott gap at half filling;
it also improves considerably the agreement with QMC results on much larger lattices, with the advantage of being free from the `minus sign problem' present in QMC data at low temperatures \cite{dosSantos03,Mondaini12}, signalled in Fig.\,\ref{fig:platx}(a) by the large error bars for $\mu\lesssim 2$.  
In this respect, it should be stressed that the corruption of QMC data gets worse in the more interesting regime, $U>W$.

The presence of a harmonic trap is taken care of through the local-density approximation (LDA).     Within LDA, we assume the fermion density a distance $r$ from the trap center is given by the uniform density [e.g., by the ABC data in Fig.\,\ref{fig:platx}(a)] at the value $\mu=\mu_0- V_tr^2$, with $\mu_0$ being determined by imposing that the \emph{total} number of fermions, $N$, is fixed: $N=2\pi\int_0^\infty n(r)r\,dr =(\pi/V_t)\int_{-\infty}^{\mu_0}d\mu\,n(\mu) $, with $n(\mu)$ being the calculated fermion density given by plots similar to Fig.\,\ref{fig:platx}(a).
In what follows, we present results obtained for $N \sim 10^3$ fermionic atoms, and typically between 50 and 80 samples of BC's; the BC averaging is also carried out over the different clusters of Fig.\,\ref{fig:clusters}.  
As discussed in Ref.\,\cite{Helmes08}, LDA provides a good description of density profiles, but can fail for spectral properties; 
see also Refs.\,\cite{Rigol04a,Duarte14}.
Though we limit our study to ground states properties, the results are relevant to actual experiments. 
Indeed, in the temperature scale $k_BT\lesssim U$, already accessible to experiments, the entropy is associated with the spin, not `charge', degrees of freedom; therefore, the density and compressibility (see below) do not vary appreciably with temperature in this regime \cite{Hart14}.  

Figure \ref{fig:platx}(b) shows typical radial distributions of atoms as the trap opening changes. 
Shallow traps allow the cloud to spread out without the atoms experiencing significant repulsion, thus favoring a single metallic phase in which the average occupancy of all sites is less than one.
As the trap narrows, phase separation sets in: the atoms confined around the trap center form an incompressible core with singly occupied sites over a radius of approximately 15 optical lattice sites; 
surrounding this Mott core, the atoms spread into a metallic state. 
Further narrowing of the trap leads to the formation of an insulating Mott ring, surrounded by metallic phases; with increasing $V_t$, the metallic core eventually becomes a `band insulating' center (i.e., the site occupancy reaches the maximum of 2 fermions per site, within a radius of only a few lattice sites).
It should be noted that when the trap opening is fixed, and the repulsion is increased, the evolution of Mott phases is reversed: metallic $\to$ ring $\to$ core.

\begin{figure}[t]
{\centering\resizebox*{7.0cm}{!}{\includegraphics*{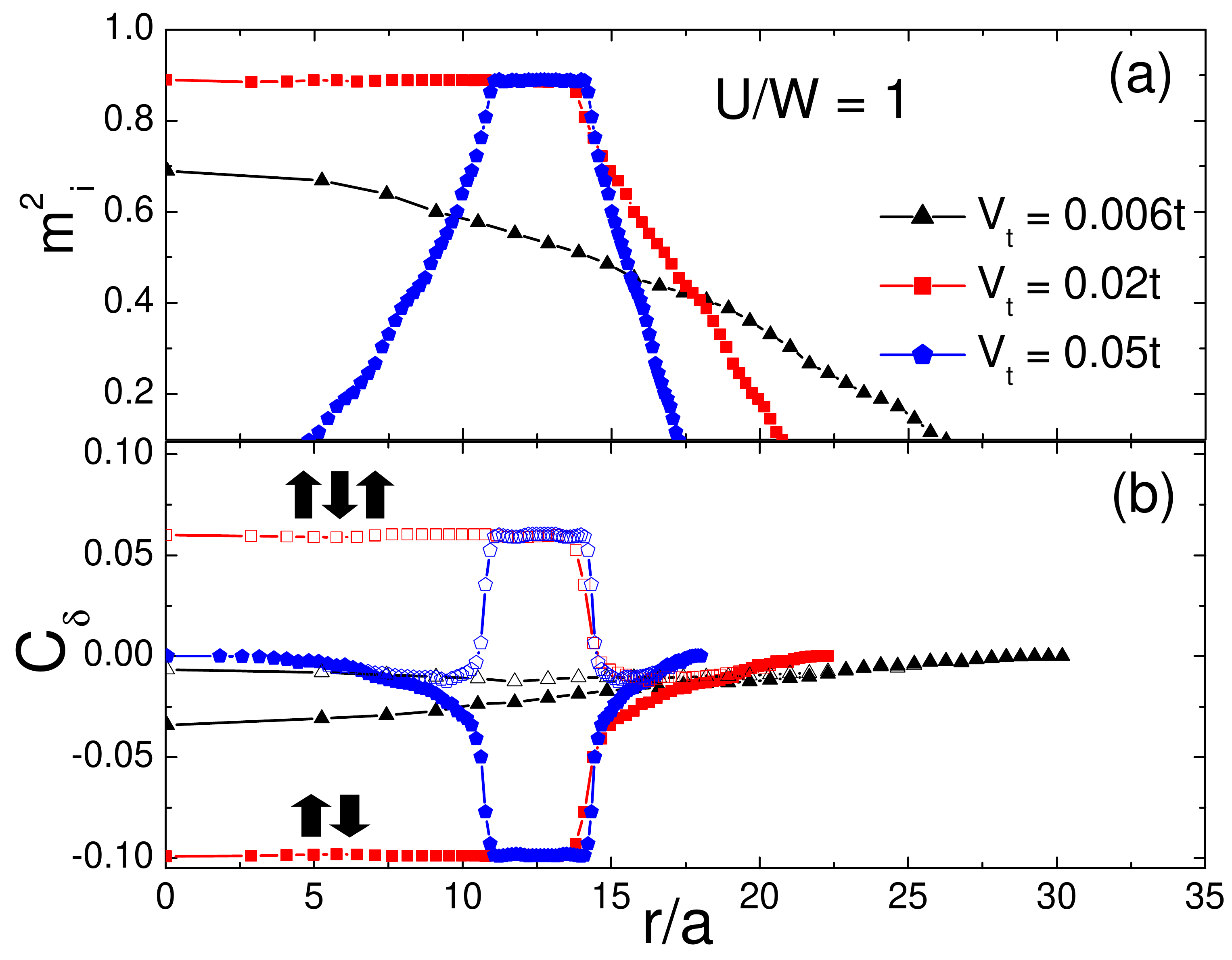}}}
\caption{(Color online) 
(a) Local moment radial distribution. 
(b) First-neighbor (filled symbols) and second-neighbor (empty symbols) spin correlation functions, calculated at different positions; the arrows just reinforce the range of correlations.
}
\label{fig:moment} 
\end{figure}

The magnetic properties of the cloud are influenced by the phase separation.
Recall that for a homogeneous system at fixed $U$, the local moment in the ground state, $m_\iv^2\equiv \ave{[n_{\iv\up}-n_{\iv\dn}]^2}$, (as a function of, say, the fermion density) is maximum at the Mott insulating state \cite{Paiva98}, though for finite $U$ it lies below the saturation value of 1, due to the non-vanishing average double-occupancy. 
Figure \ref{fig:moment}(a) shows the radial local moment distribution for the optical lattices: the nearly saturated \emph{plateaux} (for $V_t=0.02$ and 0.05) occur exactly at the half-filled sites, as depicted in Fig.\,\ref{fig:platx}(b); this provides further evidence that these rings and cores are, indeed, Mott regions.  
The presence of large local moments in the Mott regions therefore highlights where indications of long-range magnetic order should be sought. 
A second measure of magnetic ordering of these local moments is provided by the spin-spin correlation function,
$C_{\bdd}=\ave{S_\iv^zS_{\iv+\bdd}^z},$
where $\bdd=a\hat{\mathbf{x}}$, or $a\hat{\mathbf{y}}$, for nearest neighbors, and $\bdd=a\hat{\mathbf{x}}+a\hat{\mathbf{y}}$, for next-nearest neighbors; $a$ is the lattice spacing.
Figure \ref{fig:moment}(b) shows $C_{\bdd}$ for the same trap openings discussed in Fig.\,\ref{fig:moment}(a). 
Within both the Mott ring and the Mott core, the local moments are strongly correlated, that is, both nearest- and next-nearest neighbor correlations are saturated in a N\'eel arrangement; by contrast, these correlation functions are suppressed outside the Mott regions. 
For the wide trap, the correlations are much weaker, showing no saturation throughout the trap.
Therefore, in order to probe a N\'eel state, experimenters have at their disposal not just Mott cores, but Mott rings as well.

\begin{figure}[t]
{\centering\resizebox*{8.9cm}{!}{\includegraphics*{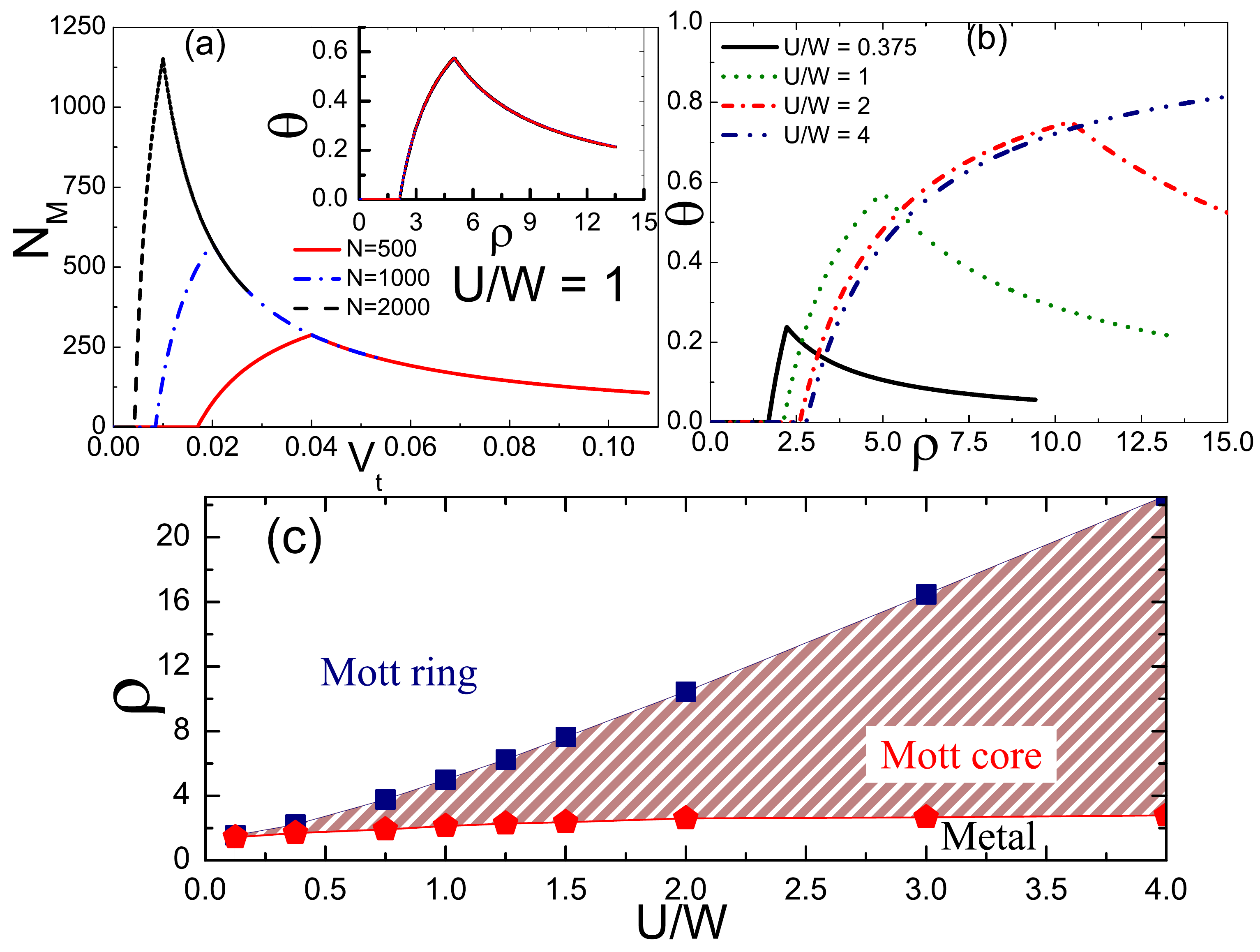}}}
\caption{(Color online) 
(a) Number of atoms in a Mott state, $N_M$, as a function of the trap depth, for different number of atoms in the cloud, $N$; the inset shows the fraction of atoms in a Mott state, $\theta\equiv N_M/N$, plotted in terms of the scaled density, which tracks $V_t$ (see text). 
(b) Mott fraction, $\theta$, as a function of the scaled density, for several values of $U$.
(c) `Phase diagram': the lower and upper curves respectively show $\rho_c(U/W)$, and $\rho_m(U/W)$ (see text).  
}
\vskip -0.6 cm
\label{fig:scaling} 
\end{figure}

This leads us to the issue of how to optimize the size of the Mott regions within the available ranges of trap parameters.
Figure \ref{fig:scaling}(a) shows the number of atoms in the Mott state, $N_M$, as a function of the trap depth, $V_t$, for a fixed value of the on-site repulsion, $U$, and for different total number of trapped atoms.  
In each case, two thresholds are clearly identified: the one at smaller $V_t$ signals the first appearance of a Mott region (the core), while the one at larger $V_t$ marks the maximum number of Mott atoms; the latter also coincides with the changeover between core and ring geometries, which is accompanied by the appearance of a metallic region in the trap center.  
Moreover, as the number of atoms increases, the Mott regions start appearing at wider traps, but with the two thresholds approaching each other. 

The presence of a trap brings about another length scale, $\zeta \equiv \sqrt{4t/V_t}$ (see Refs.\,\cite{Rigol03,Rigol04a}), which plays the role of an effective trap size.
It is therefore instructive to plot the fraction of Mott atoms, $\theta\equiv N_M/N$, in terms of an effective atomic density, $\rho=N/\zeta^2$: the data collapse into a single curve, as shown in the inset of Fig.\,\ref{fig:scaling}(a).
This perfect universal behavior (for a given $U$) can be shown to be an artifact of the LDA; nonetheless, the overall trend of Fig.\,\ref{fig:scaling}(b), which shows the evolution of the Mott coverage with $U$ can certainly be used as a guide to optimize the area of the Mott regions in actual experiments.
First, we note that the lowest threshold density, $\rho_c$, increases with $U$, though this dependence is quite weak.
By contrast, $\rho_m$, the density at which the Mott atomic fraction is peaked (above which the Mott region is a ring) displays a significant increase with $U$; likewise, from Fig.\,\ref{fig:scaling}(b) we see that the peak width also increases monotonically with $U$. 
In three-dimensional optical lattices, values of $U/W$ up to $\sim 7$ have been tuned for $^{40}$K \cite{Schneider08,Jordens08}, and up to $\sim 3$ for $^{6}$Li \cite{Hart14,Duarte14};
Fig.\,\ref{fig:scaling}(b) indicates that with values of $U/W\sim 4$ on two-dimensional optical lattices, one can expect to find at least 50\% of the atoms in a Mott state, for a wide range of effective densities.

\begin{figure}[t]
{\centering\resizebox*{8.0cm}{!}{\includegraphics*{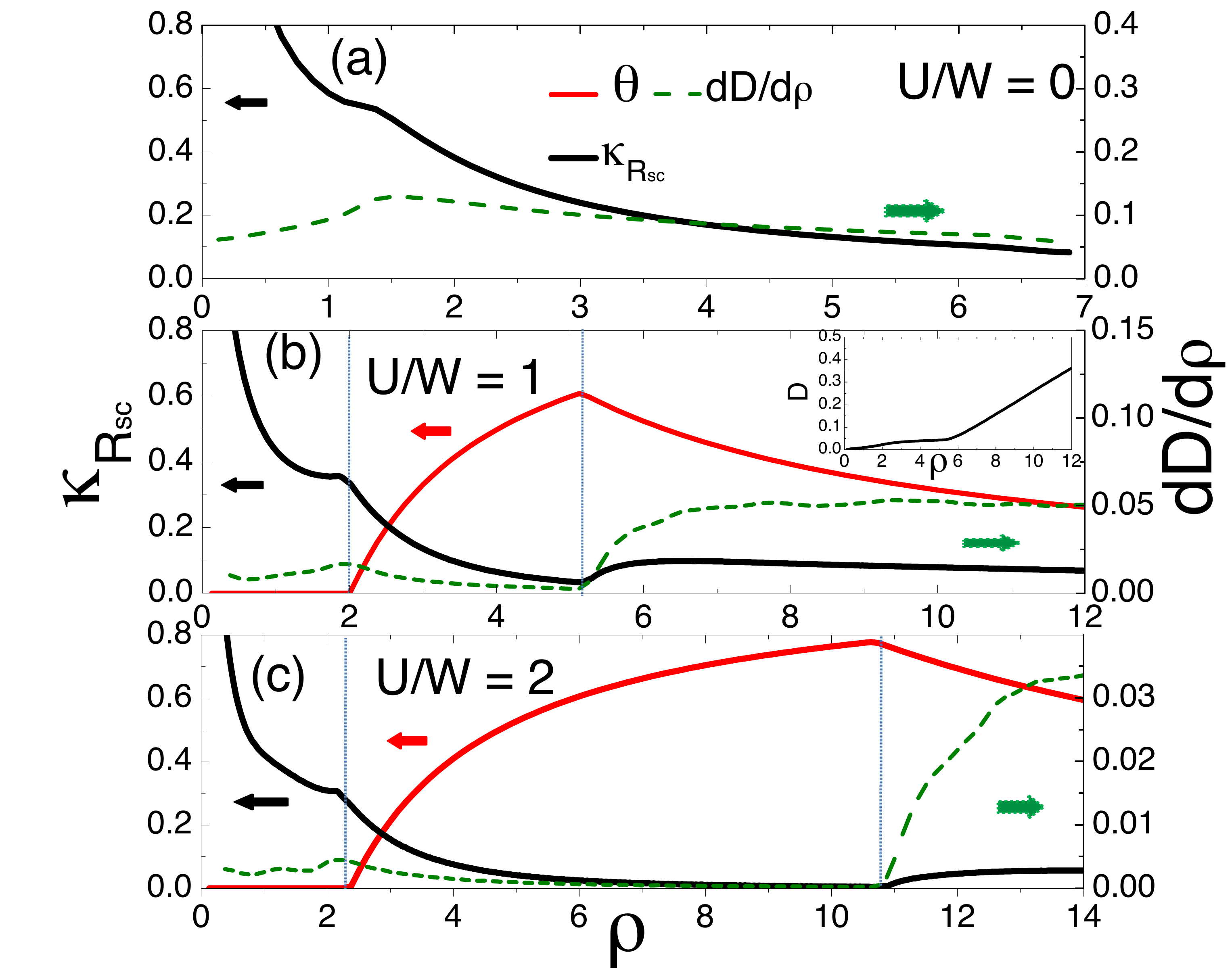}}}
\caption{(Color online) Global compressibility (full black line; left axis), derivative of double occupancy (dashed green line; right axis), and Mott coverage (full red line; left axis) as functions of the effective density. 
Each panel is for a fixed value of $U$, and the data for the compressibility have been multiplied by $\sqrt{2N}$ in order to share the left vertical axis with $\theta$. 
The inset in (b) illustrates the behavior of the double occupancy, whose derivative is plotted in the main panels. 
For $U\neq0$, the (light blue) vertical lines guide the eyes to signatures of $\rho_c$ and $\rho_m$.}
\vskip -0.5cm
\label{fig:KDr} 
\end{figure}

These results can be summarized in the form of a  `phase diagram', as in Fig.\,\ref{fig:scaling}(c): both $\rho_c$ and $\rho_m$ are shown as functions of the repulsion, $U/W$, thus highlighting the regions in parameter space in which Mott rings and Mott cores are formed. 
The diagram shows that $U$ (tuned by, e.g., a magnetic field) controls the range of effective densities in which Mott cores are found:  larger values of $U$ allow more flexibility in the choice of trap openings and number of atoms in the cloud.
It also sets limits (through $\rho_m$) beyond which one can select Mott rings instead: the larger the values of $U/W$, the tighter the traps have to be in order to generate rings.

The experimental probes used  in Ref.\,\cite{Schneider08} to detect the presence of Mott phases in the trapped cloud are the global compressibility, defined as
\begin{equation}
	\kappa_{R_{sc}}\equiv -\frac{1}{R_{sc}^2}\frac{d R_{sc}}{d\rho}, \ \text{with}\ R_{sc}=\sqrt{\frac{\ave{R^2}}{N_\sigma}},
\end{equation}
where $\sqrt{\ave{R^2}}$ is the mean-square average cloud size and $N_\sigma=N/2$ for an unpolarized gas, and the global double occupancy, 
\begin{equation}
	D=\frac{1}{N}\int d^2r\ d_{\uparrow\downarrow}(r)\,n(r), \ \text{with}\  d_{\uparrow\downarrow}=\ave{n_\uparrow n_{\downarrow}}.
\end{equation}
As we now discuss, these quantities can be used to distinguish between core and rings. 
For completeness, we mention that other definitions of compressibility, local and global, have been adapted to the context of trapped atoms \cite{Rigol03, Rigol04a,Scarola09,Taie12,Duarte14}, but they haven't been used to map shapes and sizes of the Mott regions.

 \begin{figure}[t]
{\centering\resizebox*{8.25cm}{!}{\includegraphics*{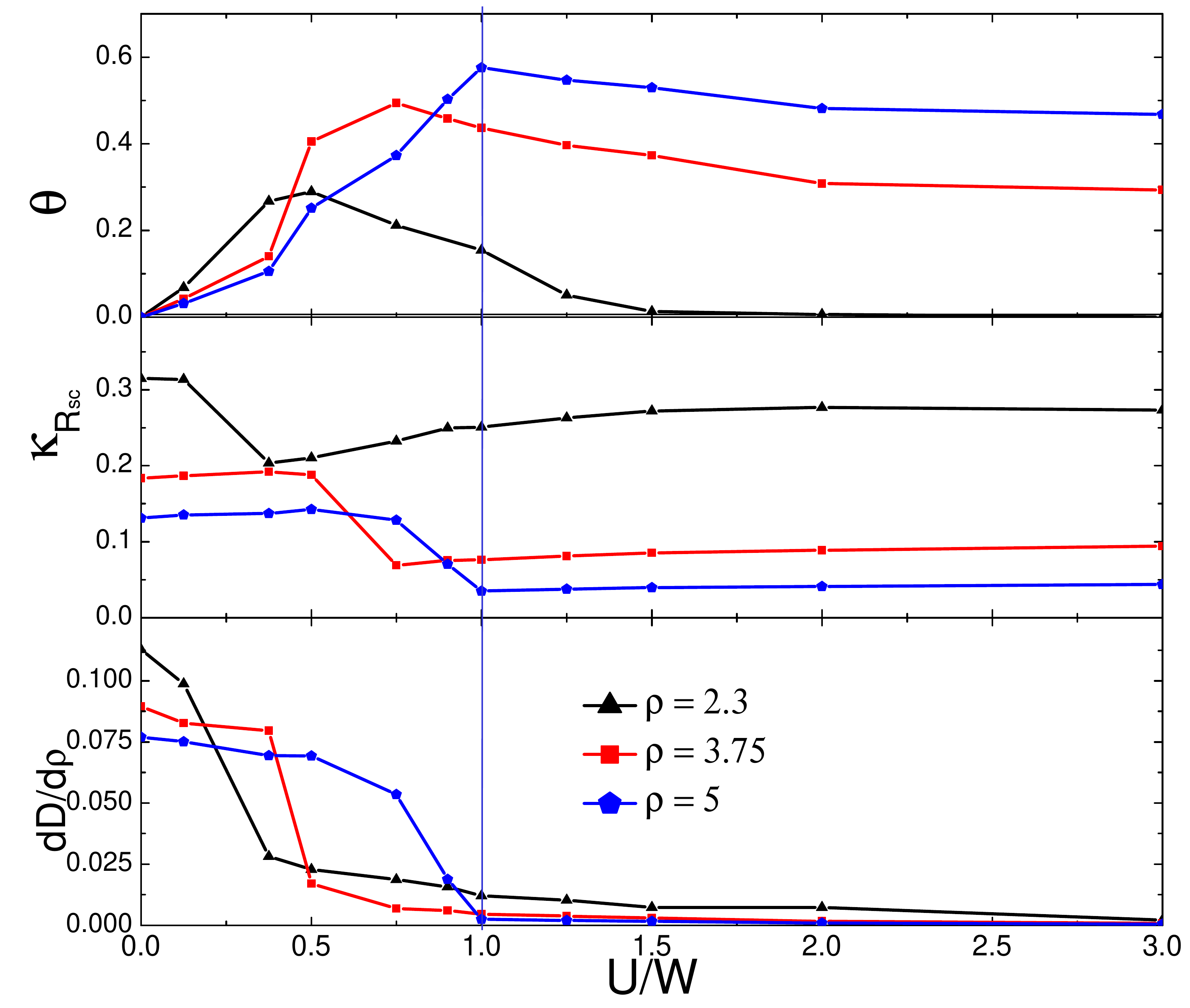}}}
\caption{(Color online) (a) Mott coverage, (b) Global compressibility, and (c) $D'$ as functions of $U/W$, for fixed effective densities.
The vertical line across the panels guides the eye through the critical value of $U/W$ for the formation of a Mott core when $\rho=5$ (see text). 
}
\vskip -0.5cm
\label{fig:KDU} 
\end{figure}

Figure \ref{fig:KDr} shows some of our calculated results for these probes; data for $\theta(\rho)$ from Fig.\,\ref{fig:scaling}(b) have also been included, for comparison.
In the non-interacting case [Fig.\,\ref{fig:KDr}(a)], the zero-temperature global compressibility $\kappa_{R_{sc}}(\rho)$ shows a hump when the chemical potential crosses the van Hove singularity; for the square lattice this occurs when sites near the trap center are singly occupied ($n=1$), while for the simple cubic lattice we determined that it occurs when this occupation is $n\sim 0.4$.
As the on-site repulsion is switched on, the hump in $\kappa_{R_{sc}}(\rho)$ persists, but is now followed by a drop; interestingly, a similar hump appears in one instance of the corresponding experimental data for the three-dimensional optical lattice \cite{Schneider08}, but, as far as we know, a systematic analysis has not been carried out.
Comparison with the plots of $\theta(\rho)$ for the same $U\neq 0$ in each panel, correlates this drop with $\rho_c$; that is, the second derivative of $R_{sc}$ appears to be discontinuous at $\rho_c$.
Further, this drop is interrupted at some larger effective density, which is now associated with $\rho_m$, beyond which $\kappa_{R_{sc}}(\rho)$ goes through a local maximum.
One should note that the latter rise was clearly visible in the experimental data for ultracold $^{40}$K atoms \cite{Schneider08}, and was actually used to signal the appearance of a Mott region in three dimensions; unfortunately, cores or `shells' could not be systematically resolved by those data, which have also been somewhat smoothed by the low, but finite, temperatures. 

As shown in the inset of Fig.\,\ref{fig:KDr}(b), the average double occupancy $D(\rho)$ rises slowly from zero for wide traps, and sharply beyond $\rho_m$: the decrease in the Mott coverage beyond this trap depth (for fixed $N$) allows double occupancy to set in between the trap center and the Mott ring; see Fig.\,\ref{fig:platx}(b).
However, a distinctive signature of $\rho_c$ cannot be unambiguously identified in the $D(\rho)$ plots.
By contrast, the main panels (for $U\neq0$) in Fig.\,\ref{fig:KDr} show that $D'\equiv \partial D/\partial \rho$ does display a local maximum at $\rho_c$, and a sudden rise at $\rho_m$.
These signatures are more distinctive for larger values of $U$, since in this range $D'$ goes to zero between $\rho_c$ and $\rho_m$, as a result of a large fraction of atoms being found in Mott states. 
Therefore, the use of $D'$ allows one to resolve these critical effective densities. 
It is also instructive to discuss the behavior with $U/W$, for three fixed effective densities, following three imaginary horizontal lines in Fig.\,\ref{fig:scaling}(c); the result is shown in Fig.\,\ref{fig:KDU}.
For $\rho=5$, the ring-core `transition' occurs at $(U/W)_c=1$, value at which a vertical line is drawn: $\theta$ is peaked, and both $\kappa_{R_{sc}}$ and $D'$ abruptly change their slopes. 
As $\rho$ decreases, $(U/W)_c$ also decreases, but for $\rho$ sufficiently small (e.g., $\rho=2.3$ in Fig.\,\ref{fig:KDU}), the Mott core cannot withstand strong repulsion and becomes metallic. 
Again, the signature of the ring-core transition can also be sought in the $U/W$-dependence of these global quantities.

In summary, by studying trapped fermionic atoms in a two-dimensional optical lattice, we have unveiled   several features, which can guide further experiments. 
A key control parameter is the effective density, $\rho\propto NV_t$ ($N$ is the number of atoms in the cloud, and $V_t$ is the trap opening, or depth), in terms of which we study the fraction and shape (i.e., cores or rings) of the Mott insulating phases. 
We have found that as the trap narrows, a Mott core forms at some $\rho_c$, which, upon further narrowing becomes a Mott ring at $\rho_m$. 
These special densities can be experimentally identified through the global compressibility, and the double occupancy (actually, its derivative with respect to $\rho$).
The importance of these special densities stems from the fact that a common measure of the size of these `phases' is the fraction of atoms in a Mott state, $\theta(\rho)$, which displays a maximum at $\rho_m$; from our calculations a 'phase diagram' for the boundaries $\rho_m(U)$ and $\rho_c(U)$ was proposed, which 
should be useful in the experimental control of the geometry of the Mott state. 
As a final comment, our findings indicate that the density of states leaves detectable traces on measurable quantities.
Hopefully, this could pave the way for studies of trapped fermionic atoms in two-dimensions addressing two long standing issues related to the Hubbard model: one is whether or not the ground state is a Fermi liquid (marginal, or otherwise \cite{Kakehashi05,Benfatto06}), and the other concerns the possibility of evolution from a Mott insulator to superconductor, as occurs in the cuprates. 
In addition, we hope our results stimulate further studies in three dimensions, e.g., establishing the behavior of the Mott volume along lines similar to the $\theta(\rho)$ we discussed here, as a step towards reaching an antiferromagnetic phase; in this respect, even in two-dimensions, though the N\'eel state is unstable at any finite temperature, imaging and controlling spin waves in square optical lattices would certainly provide new insights.

\acknowledgments 
We are  grateful to 
the Brazilian Agencies CAPES, CNPq, and FAPERJ for financial support; support from the 
INCT on Quantum Information/CNPq
is also gratefully acknowledged.

\bibliography{biblio-LDA}

\end{document}